\documentclass[12pt,preprint]{aastex}

\def\wig#1{\mathrel{\hbox{\hbox to 0pt{%
          \lower.5ex\hbox{$\sim$}\hss}\raise.4ex\hbox{$#1$}}}}

\slugcomment{Accepted for publication in ApJ, May 15 2007}

\shorttitle{}
\shortauthors{Leggett et al.}

\begin{document}

\title{Physical and Spectral Characteristics of the T8 and Later-Type Dwarfs}

\author{S. K. Leggett}
\affil{Gemini Observatory,  670 N. A'ohoku Place Hilo HI 96720}

\author{M. S. Marley}
\affil{MS 245-3, NASA Ames Research Center, Moffett Field, CA 94035}

\author{R. Freedman}
\affil{SETI, 515 N. Whisman Road, Mountain View, CA 94043}

\author{D. Saumon}
\affil{Los Alamos National Laboratory, MS F663, Los Alamos, NM 87545}
\email{dsaumon@lanl.gov}

\author{Michael C. Liu}
\affil{Alfred P. Sloan Research Fellow, Institute for Astronomy, University of Hawaii, 2680 Woodlawn Drive, Honolulu HI 96822}

\author{T. R. Geballe}
\affil{Gemini Observatory,  670 N. A'ohoku Place, Hilo HI 96720}

\author{D. A. Golimowski}
\affil{Department of Physics \& Astronomy, Johns Hopkins University, 
3400 North Charles Street, Baltimore, MD 21218}

\and

\author{D. C. Stephens}
\affil{BYU Department of Physics \& Astronomy, N-145 ESC, Provo, UT 84602}

\begin{abstract}

We use newly observed and published near--infrared spectra, 
together with synthetic spectra obtained from model atmospheres,
to derive physical properties of three of the 
latest-type T dwarfs.  A new $\rm R \approx 1700$ spectrum of the T7.5 dwarf HD~3651B, 
together with existing data, 
allows a detailed comparison to the well-studied and very similar dwarf, Gl~570D.  We find that HD~3651B 
has both higher gravity and metallicity than Gl~570D, with best-fit atmospheric parameters of $T_{\rm 
eff}=820 - 830\,\rm K$, $\log g= 5.4 - 5.5$, [m/H]$=+$0.2 and $K_{zz}=10^4$~cm$^2$s$^{-1}$.  Its  
age is 8 -- 12~Gyr and its implied mass is 60 -- 70~M$_{\rm Jupiter}$. We perform a similar analyis of the T8 and 
T7.5 dwarfs 2MASS J09393548-2448279 and 2MASS J11145133-2618235 using published data, comparing them to the 
well-studied T8, 2MASS J04151954-0935066.  We find that these two dwarfs have effectively the same 
$T_{\rm eff}$ as the reference dwarf, and similar or slightly higher gravities, but lower metallicities.
The derived parameters are $T_{\rm eff}=$725 - 775~K 
and [m/H]$=-$0.3; $\log g=5.3 - 5.45$ for 2MASS J09393548-2448279 and $\log g=5.0 - 5.3$ for 2MASS 
J11145133-261823. The age and mass are $\sim$10~Gyr and 60~M$_{\rm Jupiter}$ for 2MASS 
J09393548-2448279, and $\sim$5~Gyr and 40~M$_{\rm Jupiter}$ for 2MASS J11145133-261823. 

A serious 
limitation to such analyses is the incompleteness of the line lists for transitions of CH$_4$ and NH$_3$ at 
$\lambda \leq 1.7 \mu$m, which are also needed for synthesizing the spectrum of the later, cooler, Y type.
Spectra of Saturn and Jupiter, and of laboratory CH$_4$ and NH$_3$ gas, suggest that NH$_3$ features
in the $Y$- and $J$-bands may be useful as indicators of the next spectral type, and not  features in the $H$- and
$K$-bands as previously thought.  However, until cooler objects are found, or the line lists improve,
large uncertainties remain, as the abundance of NH$_3$ is likely to be significantly below the
chemical equilibrium value. Moreover 
inclusion of laboratory NH$_3$ opacities in our models predicts band shapes that are discrepant with 
existing data.  It is possible that the T spectral class will have to be extended to temperatures 
around 400~K, when water clouds condense in the atmosphere and dramatically change the spectral energy distribution 
of the brown dwarf.

\end{abstract}

\keywords{
infrared: stars --- stars: low-mass, brown dwarfs --- 
stars: individual (2MASS J04151954-0935066, 2MASS J09393548-2448279, 2MASS J11145133-2618235,
HD~3651B, Gl~570D)}

\section{Introduction}

The first of the T- or methane-type ultracool brown dwarfs, Gl~229B, was discovered as a companion
to an M dwarf by Nakajima et al. in 1995.  It took four more years for other examples to be found,
and these were revealed primarily as a result of  far-red and near-infrared large area sky surveys:   
the Sloan Digital Sky Survey (SDSS; York et al.\ 2000) and the Two Micron
All Sky Survey (2MASS; Beichman et al.\ 1998, Skrutskie et al.\ 2006).  There are now over 100
spectroscopically confirmed and published T dwarfs (see the on-line compendium at DwarfArchives.org).
A near-infrared classification scheme for T dwarfs, using H$_2$O and CH$_4$ absorption bands, 
has been defined by the 2MASS and SDSS groups
(Burgasser et al. 2006a).  Gl~229B is now classified as a peculiar T7
dwarf, and the other  known T dwarfs range from T0 to T8.  

In this paper we study three of the six known dwarfs later than T7 published as of February 2007. A summary of the 
astrometric properties and previously derived physical parameters of all six is given in Table 1. The uncertainty 
in spectral type classification is typically 0.5 of a subclass; hence by including T7.5 dwarfs this sample is representative 
of a ``T8'' sample. New and fainter very-late T dwarfs are now being discovered in the UKIRT Infrared Deep Sky Survey 
(UKIDSS; Hewett et al. 2006, Lawrence et al. 2007, Warren et al. 2007a).  Lodieu et al. (2007) present eight new T dwarfs, 
of which the latest is classified as a T7.5.  Warren et al. (2007b) present the latest-type T dwarf to date, a T8.5 dwarf.  
These two dwarfs are not discussed further, and the reader is directed to the discovery papers.

Three of the six dwarfs listed in Table 1, 2MASS J04151954-0935066 (T8, Burgasser et al. 2002, hereafter referred to as 
2MASS0415-09), 2MASS J12171110-0311131 (T7.5, Burgasser et al. 1999, hereafter 2MASS1217-03) and Gl~570D (T7.5, Burgasser et 
al. 2000), have been recently examined in detail by Saumon et al. (2006, 2007) using synthetic spectra calculated from model 
atmospheres, and ground- and space-based observational data covering the wavelength range 0.9---15~$\mu$m. Saumon et al. 
(2007) show that 2MASS1217-03 is more metal-rich and hotter than the other two T dwarfs, with effective temperature $T_{\rm 
eff}\approx$900~K.  Gl~570D has an effective temperature of $T_{\rm eff}\approx 800\,\rm K$ and 2MASS0415-09 has a lower 
$T_{\rm eff}$ of around 750~K. In this paper we use Gl~570D and 2MASS0415-09 as reference objects for the other 
three dwarfs listed in Table 1, 2MASS J09393548-2448279 and 2MASS J11145133-2618235 (Tinney et al. 2005, hereafter 
2MASS0939-24 and 2MASS1114-26), and HD~3651B (Murgauer et al. 2006, Luhman et al. 2007).

Burgasser, Burrows \& Kirkpatrick (2006b, hereafter BBK) present a method for determining the physical parameters of the 
coldest T dwarfs using low-resolution near-infrared spectra. BBK use an index measuring the strength of the H$_2$O feature 
at 1.15~$\mu$m, H$_2$O$-J$, as a measure of $T_{\rm eff}$, while the ratio of the 2.1~$\mu$m flux to the 1.6~$\mu$m flux, 
$K/H$, is used to measure gravity. They use modelled values of these ratios, calibrated by the $T_{\rm eff}$ and gravity of 
the well-studied dwarf Gl~570D, to determine $T_{\rm eff}$ and gravity for a sample of T6 to T8 dwarfs.  The values for the 
dwarfs in our sample are given in Table 1.  BBK suggest that 2MASS0939-24 and 2MASS1114-26 are the coolest 
dwarfs known, with $T_{\rm eff}$s cooler than that of 2MASS0415-09 by $\gtrsim$50~K. However, the gravity index of BBK is also 
sensitive to metallicity, as shown for example by Liebert \& Burgasser (2007) and Liu, Leggett \& Chiu (2007, hereafter 
LLC).  This could affect the conclusions drawn by BBK; thus here we re-examine the properties of these two very-late T 
dwarfs.

We also examine in detail the atmospheric properties of HD~3651B.  Luhman et al. (2007), Burgasser (2007) and LLC use 
near-infrared spectra and (for Luhman et al.) IRAC 3--8~$\mu$m photometry, to determine the physical parameters of 
this dwarf. The values of LLC and BBK are given in Table 1 (the LLC values agree well with those of Luhman et al.).  
This  dwarf is of particular interest, as its metallicity and age can be constrained from its primary star, which is only 
possible for  one other of the very late T dwarfs, Gl~570D. HD~3651B is also of interest as it is a wide companion to a star 
which hosts a planet (Fischer et al. 2003).

Finally, as several groups continue to search for dwarfs later and cooler than these (e.g. UKIDSS), we also explore what 
such objects may look like, using the observed spectra of the solar system gas giant planets Jupiter and 
Saturn, and laboratory spectra of CH$_4$ and NH$_3$ gas.

\section{Known Properties of the Sample}

\subsection{The K~V $+$ T7.5 Systems: Gl~570 and HD~3651}

Figure 1 shows previously published low-resolution near-infrared spectra for the T7.5 dwarfs
Gl~570D and HD~3651B. It can be seen that the spectra are quite similar, with HD~3651B being
slightly brighter in the 1.6~$\mu$m $H$-band and more significantly brighter in the 2.1~$\mu$m
$K$-band.

The Gl~570 system consists of a K dwarf, two early-M dwarfs, and the T dwarf.  The last was discovered in the 2MASS database 
by Burgasser et al. (2000) as a very widely separated 258$\farcs$3 (1525~AU projected separation) companion to the system.  
HD~3651B is also a companion to a K dwarf, at a distance of 42$\farcs$9 (476~AU projected separation), and the 
primary is known to be an exoplanet host. HD~3651B was discovered in near- and mid-infrared imaging searches for companions 
to exoplanet stars by Mugrauer et al. (2006) and Luhman et al. (2007).  The metallicities of both primaries are known: 
Gl~570A has [m/H]$=$ +0.00 to +0.07 (Feltzing \& Gustafsson 1998, Santos et al. 2005), and HD~3651A has [m/H]$=$ +0.09 to 
+0.16 (Gray et al. 2003, Santos et al. 2004, Valenti \& Fischer 2005).

Evolutionary models, constrained by the observed luminosity and further refined by comparison to synthetic 
spectra, show that the age of Gl~570D is 3 -- 5~Gyr, with $T_{\rm eff}= 800 - 820\,\rm K$, $\log g= 5.09 - 
5.23$ and mass $38 - 47\,\rm M_{Jupiter}$, where the lower limit to $T_{\rm eff}$, log~$g$ and mass 
correspond to the youngest age (Saumon et al. 2006).  Similar analysis shows that the age of HD~3651B is 3 
-- 12~Gyr, with $T_{\rm eff}= 780-840\,\rm K$, $\log g= 5.1-5.5$ and mass $40-72\,\rm M_{Jupiter}$, where 
the younger age gives lower $T_{\rm eff}$, log~$g$ and mass (LLC).  However, LLC show that the age of the 
HD~3651 system {\it relative} to the Gl~570 system can also be determined.  The log($R'_{HK}$) 
chromospheric activity parameters suggest that the HD~3651 system is about five times older than the 
Gl~570 system; the X-ray luminosities suggest that the HD~3651 system is about twice as old as the Gl~570 
system; and isochrone analyses suggest that the HD~3651 system is at least 2.5 times older than the 
Gl~570 system.  Hence if the age of Gl~570D is 3 -- 5~Gyr, then the age of HD~3651B is 8 -- 12~Gyr and the dwarf 
has $T_{\rm eff}=$ 825 -- 840, log~$g=$ 5.4 -- 5.5 and mass 60 -- 72~M$_{Jupiter}$ (Figure 3 of LLC).

A higher value of $\log g$ for HD~3651B than Gl~570D would seem discrepant 
with the observed brighter $K$-band flux peak of HD~3651B (Figure 1),
as a higher gravity leads to stronger $K$-band H$_2$ absorption at this wavelength (\S 3).
Because HD~3651B is brighter than Gl~570D at $K$, 
Burgasser (2007), applying the BBK technique  to derive gravity, determines
a lower gravity for HD~3651B of $\log g= 4.7$, if the HD~3651 system has solar metallicity. 
However, using atmospheric models, Burgasser shows that increasing metallicity results in a
higher indices-implied gravity, and he derives values that agree with the luminosity- and age-driven
LLC values if the system has [m/H]$=+$0.24.

\subsection{2MASS0415-09, 2MASS0939-24 and 2MASS1114-26}

Figure 2 shows previously published low-resolution near-infrared spectra for the 
T8 dwarfs 2MASS0415-09 and 2MASS0939-24, and the T7.5 dwarf 2MASS1114-26.
It can be seen that the spectra are similar, but 2MASS0939-24 and 2MASS1114-26 
have broader 1.0~$\mu$m $Y$-peaks and are fainter at $K$ than 2MASS0415-09.

These three dwarfs  were all discovered as
isolated field T dwarfs in the 2MASS survey (Burgasser et al. 2002, Tinney et al. 2005).
Their ages  cannot be well constrained; however, the measured or estimated tangential 
velocities (see Table 1) suggest that they are not likely to be either very young, or members of the halo
(see for example the discussion of kinematics in the solar region by Eggen (1998), and references therein).
Their kinematics suggest ages of 1 -- 10~Gyr.  
Evolutionary models, constrained by observed luminosity and comparisons with model spectra, 
show that the age of 2MASS0415-09 is 3 -- 10~Gyr, with
$T_{\rm eff}=$ 725 -- 775~K and log~$g=$ 5.00 -- 5.37, where the younger age corresponds to lower $T_{\rm eff}$
and  log~$g$ (Saumon et al. 2007).  Metallicity could not be well constrained by Saumon et al.
as both [m/H]=0 and +0.3  models gave good fits to the data. 

The spectral classification indices for all three of these dwarfs are very similar (Burgasser et al. 2006a), despite the 
fact that 2MASS1114-26 is classified 0.5 of a sub-type earlier. The H$_2$O$-J$- and 
H$_2$O$-H$ indices are essentially identical, while the CH$_4-J$, CH$_4-H$ and CH$_4-K$ indices place 
2MASS0939-24 and 2MASS1114-26 at a very slightly earlier type than 2MASS0415-09 (see Figure 2).

\section{Model Atmospheres}

\subsection{Description of the Models}

Models of ultracool-dwarf atmospheres have been developed by members
of our team (Ackerman \& Marley 2001; Marley et al.\ 2002; Saumon et al.\ 2003). For this work,
a grid of models and synthetic spectra have been calculated  with
effective temperature $T_{\rm eff}=600$ -- 850~K in steps of 50~K; gravity $g=$300, 1000 and
3000 m/s$^2$ (or in the commonly-used c.g.s units, log~$g=$4.5, 5.0 and 5.5); and metallicity
[m/H]$=-0.3$, 0 and $+0.3$.  To begin to explore lower temperatures, 500~K and 550~K solar metallicity models
with $g=$300 and 1000 m/s$^2$ have also been calculated.

These models yield temperature-pressure-composition structures under conditions of 
radiative-convective equilibrium using 
the thermal radiative transfer source function technique of Toon et al.\ (1989).  For the very low-temperature dwarfs 
considered here, cloud decks of iron and silicate condensates are expected to lie below the photosphere and do not impact 
the spectral energy distribution except by reduction of the abundances of O, 
Fe, Si, Mg, Ti, and other refractory elements.  Thus we neglect grain opacities in the atmospheric structure and spectral 
calculations, although grain formation is included in the calculation of the chemical equilibrium.

The chemical equilibrium calculations for the atmosphere models are described in Lodders \& Fegley (2002) 
and use the solar system  
abundances of Lodders (2003).  There is debate about the solar CNO abundance values (e.g. Ayres, Plymate \& Keller 2006, 
Allende Prieto 2007), which strongly impact brown dwarfs' spectral energy distributions (SEDs) as they are shaped by 
opacities involving these species. Studies of line formation in the solar atmosphere by Asplund et al. (2004) have 
determined carbon, oxygen and nitrogen abundances 0.13--0.17~dex (35--50\%) 
lower than those previously given by Grevesse \& 
Sauval (1998).  The Lodders (2003) CNO abundances are 0.07--0.12~dex higher than the Asplund et al. values, 
and 0.02--0.07 dex  lower than the Grevesse \& Sauval (1998) values. The total mass fraction of all heavy elements is 
0.015 compared to Grevesse \& Sauval's 0.017 and Asplund et al.'s 0.012.

The gas opacity database includes the molecular lines of H$_2$O, CH$_4$, CO, NH$_3$, H$_2$S, PH$_3$, TiO, VO, CrH, FeH and 
CO$_2$, the atomic lines of the alkali metals (Li, Na, K, Rb and Cs), and continuum opacity from H$_2$ collisionally induced 
absorption (CIA), H$_2$, H and He Rayleigh scattering, bound-free opacity from H$^-$ and H$_2^+$, and free-free opacity from 
H$^-$, He, H$_2^-$, and H$_2^+$ (Freedman, 
Marley \& Lodders 2007).  For  further discussion of the opacities important in brown dwarf photospheres see Sharp \& 
Burrows (2007). For this work, the line list for CH$_4$ (which does not include transitions at $\lambda < 1.5 \mu$m) has been 
supplemented with the Strong et al. (1993) CH$_4$ laboratory 
data for 1.05--1.5~$\mu$m and the CH$_4$ opacity derived from the spectrum of Saturn by Karkoschka (1994) below 1$\,\mu$m.
The line list for NH$_3$ does not include wavelengths shorter than 1.4~$\mu$m, and is incomplete at 1.4--1.9~$\mu$m. We 
discuss this further in \S 7.

The models have been further advanced to allow for the inclusion of turbulent vertical
mixing, which draws up long-lived chemical species from deep, hot layers into
the cool radiative photosphere, causing their observed abundances to be greater than expected for the case of
chemical equilibrium.  The mixing is parameterized using a diffusion coefficient 
$K_{zz}$. In the turbulent convection zone we use mixing length theory to
derive $K_{zz}$, but we treat it as a tunable parameter in the nominally stable radiative zone. 
The abundances of very stable species like
CO and N$_2$ are enhanced in the upper atmosphere relative to their
chemical-equilibrium values, while those of H$_2$O, CH$_4$ and NH$_3$
are reduced (see e.g. discussion in Saumon et al. 2007 and references therein).
For T dwarfs, non-equilibrium chemistry is very important at mid-infrared wavelengths as
the abundances of CO and NH$_3$ can be affected by an order of magnitude or more
(e.g. Figure 3 of Saumon et al. 2006),
and these species have strong features around 5~$\mu$m and 10~$\mu$m. 
The effect is not as important in the near-infrared  as the abundances of species producing
the majority of the opacity there, H$_2$O and CH$_4$, are not as strongly impacted.

\subsection{Comparison of Synthetic and Observed Spectra}

Figure 3 shows synthetic spectra calculated from model atmospheres with a range of $T_{\rm eff}$, log~$g$ and metallicity 
appropriate for the dwarfs studied here. Vertical mixing is not included in these spectra as the effects in the 
near-infrared are insignificant (see Figure 2 of Saumon et al. 2006). Temperature, gravity and metallicity do however 
impact the near-infrared spectral energy distribution. The CH$_4$ and H$_2$O absorption bands are so strong at these 
temperatures that the effect of changing $T_{\rm eff}$ is quite subtle; the largest effects are calculated to occur in the 
blue and red wings of the $H$-band flux peak, and at the peak of the $K$-band. Increasing metallicity 
has a similar effect on the $K$-band
to decreasing gravity, although the effect on the $Y$-band differs.  The 2~$\mu$m flux is very sensitive to 
gravity and metallicity because of the pressure-induced H$_2$ opacity; the 1~$\mu$m flux is controlled by the H$_2$O opacity 
which is more sensitive to metallicity than gravity.

Modelling these atmospheres is extremely complex. While great advances have been made (and continue to be made), there are 
known deficiencies and uncertainties in the models.  In particular, the molecular line lists for CH$_4$ and NH$_3$ are 
incomplete or even nonexistent at $\lambda\leq$1.7~$\mu$m (e.g. Saumon et al. 2007). Figure 4 tests the models by comparing 
the synthetic and observed spectra for the two well-studied dwarfs 2MASS0415-09 and Gl~570D.  Note that the synthetic 
spectra have not been scaled to fit, but instead the known distance and structural calculations of the radius (determined by 
$T_{\rm eff}$ and gravity) are combined to determine the flux at the Earth for each dwarf. The observed luminosity and 
robust evolutionary calculations can constrain $T_{\rm eff}$ and gravity for the dwarfs to a small allowed range as 
described in \S 2. The synthetic spectra in Figure 4 fall within these ranges although they may not be optimal at the 0.2 
dex level for gravity and metallicity.

Figure 4 illustrates the successes and remaining failures of current models. Given the complexity of the
model physics, the match to the observations is  good, but there are known problems.
Incompleteness in the CH$_4$ and NH$_3$ 
opacities most likely lead to the excess model flux in the $Y$, $J$ and $H$ bands (this is discussed further in \S 7).  The 
K~I doublet at 1.25~$\mu$m is too strong; this is probably due to remaining errors in the temperature-pressure profile for 
the atmosphere, to which the condensation of K~I into KCl is very sensitive (e.g. Lodders 1999 Figure 2).  The shortfall in 
flux in the $K$-band may be due to low CNO abundances (mimicking metal deficiency), or to errors in the 
pressure-induced H$_2$ opacity.

Using Gl~570D and 2MASS0415-09 as templates, and understanding the remaining model deficiencies, we now study the three 
very-late T dwarfs, HD~3651B, 2MASS0939-24 and 2MASS1114-26 in detail.  We start with HD~3651B, for which new data have been 
obtained.

\section{New Observations of HD~3651B}

The Gemini Near-Infrared Spectrograph (GNIRS, Elias et al. 2006) was used in cross-dispersed mode with a 
0$\farcs$3 slit 
to obtain a 0.9 -- 2.5~$\mu$m R=1700 spectrum of HD3651B on 2006 September 23, through program GS-2006B-DD-5.  
The exposure time was 600~s, and HD3651B was nodded $6^{\prime\prime}$ along the slit in one ``ABBA'' pattern, for a total 
of 40 minutes on-source integration time. The F2V star HIP 118301 was used to remove telluric features, and 
flat fields and wavelength calibration 
were obtained using lamps in the on-telescope calibration unit.  The data were reduced using IRAF
and FIGARO software packages.  A slit position angle of 200 degrees was used, and no contamination 
was seen from the bright primary $43^{\prime\prime}$ to the north-west.

Figure 5 shows the final spectrum, compared to a similar resolution spectrum from McLean et al. (2003) for Gl~570D.  As  
in the low-resolution spectra for these dwarfs (Figure 1), the two spectra are very similar, and what looks like noise is in 
fact real structure due to, primarily, H$_2$O and CH$_4$ absorption.  The apparent feature seen in Figure 5 for HD~3651B 
around 1.27~$\mu$m is not seen in Figure 1 and is
most likely due to incomplete removal of the strong sky lines  at this wavelength.
The only significant difference between 
HD~3651B and Gl~570D appears in the $K$-band, where the absorption features of HD~3651B are stronger.

\section{Re-Examination of HD~3651B}

\subsection{Near-Infrared Spectra}
 
Figure 6 shows the observed low-resolution near-infrared spectrum of HD~3651B and compares it to synthetic 
spectra with $T_{\rm eff}=$ 800 and 850~K, log~$g=$ 5.0 and 5.5, and metallicity [m/H]$=$ 0 and +0.3,
covering the range of values determined for this dwarf by LLC (\S 2.1). 
These spectra have been scaled to the flux at the Earth, given the known distance to the dwarf and the 
calculated radii appropriate for the values of $T_{\rm eff}$, $g$ and [m/H]. Comparison of Figure 6 to the 
template model fits of Figure 4 shows that for either the 800~K or 850~K models, the log~$g=$ 5.5 and 
[m/H]$=$ 0 spectra produce too low a $K$-band flux, and hence this combination of gravity and metallicity 
can be excluded. An average of the log~$g=$5.0 and 5.5, [m/H]$=$ 0 and $+0.3$, 800~K spectra shown in the 
top panel, weighted towards the higher gravity spectrum for less flux at 1.2~$\mu$m, would produce a 
similar fit to the one seen for Gl~570D in Figure 4.  However the flux at the $H$ and $K$ peaks would be 
low, implying a slightly hotter temperature as suggested by the 850~K log~$g=$5.5 [m/H]$=+0.3$ comparison. 
A temperature as high as 850~K can be excluded by the excessive 1.60 -- 1.75~$\mu$m flux.  

The bottom panel compares the observed spectrum to a synthetic spectrum for $T_{\rm eff}=$ 825~K, 
log~$g=$5.5 and [m/H]$=+0.3$ (calculated by averaging the 800 and 850~K spectra). The
quality of the fit is indeed similar to that for Gl~570D in Figure 4, although a slightly lower 
metallicity would improve the match at $Y$, and this would then require a decrease in 
gravity to maintain the match at $K$.   Hence our 
best  fits have $T_{\rm eff}\approx$ 825 with log~$g=$5.4 -- 5.5 and  [m/H]$\approx +0.2$.
This metallicity is somewhat higher than the measured values of $+0.09$ to 
$+$0.16 (Gray et al. 2003, Santos et al. 2004, Valenti \& Fischer 2005).  However Valenti \& Fischer find 
a range for individual elements from $+$0.11 for [Ti/H] to $+$0.24 for [Na/H] and so [m/H]$\approx +0.2$ 
is not unreasonable.  Note that although there are uncertainties in the absolute values of the solar 
chemical abundances (\S 3.2), they do not affect the conclusion reached here that HD~3651B is 0.2 dex 
more metal-rich than Gl~570D, which has effectively solar chemistry (\S 2.1).

Figure 7 plots the observed medium-resolution $K$-band spectra of Gl~570D and HD~3651B with the synthetic 
spectra for the [$T_{\rm eff}$, log~$g$, [m/H]] models [800, 5.0, 0.0] and [800, 5.5, +0.3]. Although 
there are discrepancies between the data and the models, 
the relative depths of the observed and modelled features around 
2.05~$\mu$m and 2.14 -- 2.18~$\mu$m (see also Figure 5) supports the conclusions reached above that HD~3651B has 
a higher gravity and higher metallicity than Gl~570D.

\subsection{IRAC Photometry}

$Spitzer$ IRAC mid-infrared four-channel photometry (Fazio et al. 2004) is available for both Gl~570D 
(Patten et al. 2006) and HD~3651B (Luhman et al. 2007).  Saumon et al. (2006) show that the 6 -- 14~$\mu$m 
spectrum of Gl~570D obtained with the $Spitzer$ Infrared Spectrograph (IRS, Houck et al. 2004) matches 
very well the synthetic spectrum generated from a model with $T_{\rm eff}=$821~K, log~$g=$5.23, 
[m/H]$=$0.0 and $K_{zz}=10^2$~cm$^2$s$^{-1}$.  As the distances to the two dwarfs are known, we can 
compare the IRAC absolute magnitudes calculated by the models to the observed values (note that the IRAC 
magnitudes cannot be literally interpreted as a flux at the nominal filter wavelength, Cushing et al. 
2006).

Figure 8 plots the observed IRAC absolute magnitudes for the two dwarfs, together with modelled values for 
$T_{\rm eff}=$825~K and vertical mixing coefficients $K_{zz}=0$ (i.e. equilibrium chemistry) and 
$K_{zz}=10^4$. The modelled values have been determined by interpolating between the 800 and 850~K 
synthesized photometry. Log~$g=5.25$ and [m/H]$=0$ values appropriate for Gl~570D are shown, as well as 
log~$g=5.5$ and [m/H]$=+0.3$ values, appropriate for HD~3651B. The agreement between observed and modelled 
values for both dwarfs is excellent for the non-equilibrium chemistry case, with an average deviation of 
0.10 magnitudes. The deviation between the observed and modelled $K_{zz}=0$ magnitudes is on average 0.2 
magnitudes for both dwarfs, with individual band discrepancies as large as 0.3 magnitudes. Hence 
non-equilibrium chemistry must be included in studies of the mid-infrared fluxes of the late T dwarfs, as 
also found by Golimowski et al. (2004), Patten et al. (2006) and Leggett et al. (2007).

\subsection{Adopted Parameters for HD~3651B}

Our best-fit parameters for HD~3651B are summarised in Table 2. 
Comparison to the well-studied dwarf Gl~570D, and synthetic spectra, show that 
HD~3651B has $T_{\rm eff}=$820 -- 830~K, log~$g=$5.4 -- 5.5 and [m/H]$\approx +0.2$.
The corresponding age and mass is 8 -- 12~Gyr and 60 -- 70~M$_{\rm Jupiter}$
(Figure 3 of LLC).
Our derived gravity agrees with the high end of the 
range determined by LLC (Table 1), which is consistent with an age at the high end of their allowed 
range. The latter is in turn consistent with the relative ages of the Gl~570 and HD~3651 systems (\S 2.1).  
The low end of our range of values for $T_{\rm eff}$,  log~$g$, age and mass agree with the
values derived by Burgasser (2007) when the BBK flux indices are recalibrated for a metallicity
of [m/H]$=+$0.24, consistent with the value implied by our model fit.  The Burgasser values for
lower metallicity require the HD~3651 system to be younger than 4.7~Gyr (see his Table 1), 
which is inconsistent with the various age indicators described in \S 2.1.

\section{Re-examination of 2MASS0939-24 and 2MASS1114-26}

The distances to 2MASS0939-24 and 2MASS1114-26 have not yet been published and so an absolute comparison of synthetic to 
observed spectra, i.e. with a known scale factor similar to Figures 4 and 6, cannot be made for these dwarfs.  Neither can 
their luminosities be measured and $T_{\rm eff}$ derived from structural models, as has been done for 2MASS0415-09, Gl~570D 
and HD~3651B. Nevertheless, a comparison of the model spectra shown in Figure 3 to the observed spectra shown in Figure 2 
allows us to constrain $T_{\rm eff}$, log~$g$ and [m/H] for them, relative to the well-studied 
dwarf 2MASS0415-09. As described in \S 2, Saumon et al. (2007) shows that the age of 2MASS0415-09 is 3 -- 10~Gyr, with 
$T_{\rm eff}=$ 725 -- 775~K and log~$g=$ 5.00 -- 5.37; metallicity could 
not be constrained. We suggest that 2MASS0415-09 has solar or only slightly higher metallicity, as Figure 
4 shows that the quality of the fit, especially in the metallicity-sensitive $K$-band, is very similar to that of Gl~570D, 
for which a solar metallicity has been measured (\S 2.1).

Figure 3 shows that the ratio of the $H$-band flux peak to the depth of the CH$_4$ absorption around 1.7~$\mu$m, is 
sensitive to $T_{\rm eff}$.  Although the model opacities are incomplete in this region, we assume the strength of this 
absorption can still be used for relative estimates of $T_{\rm eff}$.  Hence $T_{\rm eff}$ for 2MASS0939-24 and 2MASS1114-26 
appears to be essentially identical to that for 2MASS0415-09, as would also be surmised from the strengths of the H$_2$O 
absorptions at 1.15 and 1.45~$\mu$m.

The suppressed $K$-band flux for 2MASS0939-24 and 2MASS1114-26, relative to 2MASS0415-09,
 can be explained by either higher gravity or lower metallicity. These effects however can be distinguished by their effects 
on the $Y$-band, as shown in Figure 3. (Again, although the models do not perfectly reproduce the $Y$-band observation, we 
believe that comparisons in a relative sense are valid.) The relative heights of the $K$-bands and the width of the $Y$-band 
flux peaks suggest that both 2MASS0939-24 and 2MASS1114-26 are quite metal-poor compared to 2MASS0415-09. Adopting solar 
metallicity for 2MASS0415-09 implies [m/H]$\approx-0.3$ for these two dwarfs.  The additional suppression at $K$ for 
2MASS0939-24 suggests that it also has a higher gravity than 2MASS0415-09.

Our adopted parameter ranges for these dwarfs are given in Table 2 and are: $T_{\rm eff}=$ 725 -- 775~K, [m/H]$\approx-0.3$, and 
log~$g\approx$5.0 -- 5.3 for 2MASS1114-26 and 5.3 -- 5.45 for 2MASS0939-24.  The upper gravity limit for 2MASS0939-24 is 
determined by evolutionary arguments, assuming an upper age limit of 12~Gyr for the dwarf. We do not confirm the lower 
temperature of $\lesssim$700~K determined for these dwarfs by BBK (Table 1); instead we find that these dwarfs each have 
$T_{\rm eff}$ essentially identical to that of 2MASS0415-09, and suggest that the method of BBK needs to be revised to
take metallicity into account (as done for HD~3651B by Burgasser 2007).

Saumon et al. (2007) and LLC show that $T_{\rm eff}$ and gravity give mass and age for a brown dwarf, and 
that changing metallicity by 0.3 dex impacts derived mass and age by $<$5\%.  We derive ages and 
masses for these metal-poor late-T dwarfs of 10 -- 12~Gyr and 50 -- 65~M$_{\rm Jupiter}$ for 
2MASS0939-24, and 3 -- 8~Gyr and 30 -- 50~M$_{\rm Jupiter}$ for 2MASS1114-26.  When available, parallaxes 
will constrain these parameters further.

\section{Dwarfs Cooler than 700~K and the Next Spectral Type}

\subsection{Expected Spectral Trends and the Opacity Problem}

Many researchers are currently searching for objects cooler than the coldest known brown dwarfs with $T_{\rm 
eff}\approx$700~K, and in particular for the next, cooler, spectral type, for which the letter ``Y'' has been suggested 
(e.g. Kirkpatrick 2005).  Until examples of cooler dwarfs are found one cannot be certain what spectral change will
trigger the use of a new letter; if no obvious spectral changes are seen, the T classification will need to 
be extended beyond T8.  

For practical reasons, the spectral indicators for classification of the extreme-T or Y dwarfs should 
remain in the far-red or near-infrared, and should be identifiable
based on low spectral-resolution data, as these dwarfs will be faint. 
As the temperature decreases
more flux emerges at mid-infrared wavelengths, relative to the red and near-infrared; 
the calculated ratios of the 4.5~$\mu$m and 10.0~$\mu$m
fluxes relative to the $J$-band 1.27~$\mu$m flux, in Jy, are 2.3 and 1.2 
at 600~K, and 3.4 and 1.7 at 500~K.  
However, detector sensitivity in the mid-infrared is worse than in the 
near-infrared, such that, 
for example, it would take 2--5 times longer to obtain a  5~$\mu$m $Spitzer$ spectrum than a 
$J$-band spectrum at the same S/N ratio with current 
instrumentation on 8~m-class telescopes.  Although $JWST$ is expected to be very sensitive in the 
mid-infrared, it will be $\sim$6 magnitudes more sensitive in the near-infrared. 

In this Section we use planetary and laboratory data,  and new model calculations, to 
investigate spectral trends in the near-infrared as $T_{\rm eff}$ drops 
below 700~K. If the changes are too subtle to trigger a new spectral type, it will still be important 
to find and understand indicators of lower temperature in the 1.0--2.2~$\mu$m region,
as very little flux remains in the H$_2$O and CH$_4$ absorption
features currently used to classify the T dwarfs (see Figures 2 and 3).

A summary of the expected spectral changes at $T_{\rm eff}<$~700~K is given by Burrows, Sudarsky \& Lunine (2003).  
They describe how, around $T_{\rm eff}=$600~K, the strong red and near-infrared absorption features of K~I weaken 
as this element forms grains of KCl, and NH$_3$ features become apparent in the blue wings of the $H$ and $K$ flux 
peaks, around 1.5 and 1.95~$\mu$m.  Below 400~K, water clouds are expected to form causing 
the colors of dwarfs
to change dramatically. It is possible that the NH$_3$ features
in the $H$- and $K$-bands may trigger the next spectral type. 
(NH$_3$ is already seen at 10~$\mu$m in T dwarfs (Roellig et al. 2004),
however its appearance in the more easily measured near-infrared could be used as a trigger 
in the same way that CH$_4$ at $H$ and $K$ is used to trigger the T type, 
despite CH$_4$ absorption being prominent at 3~$\mu$m in mid- and late-L dwarfs (Noll et al. 2000)).
However, although our 700~K and 600~K model spectra in Figure 3 (and our 500~K  spectra, not shown)
show a change in the shape of the $Y$-band flux peak, and a steady 
decline of the $K$-band flux, there is no obvious spectral change that would indicate the necessity of a new 
spectral type. 

Unfortunately, both our analysis and that of Burrows, Sudarsky \& Lunine, suffer from incomplete or non-existent
line lists for the important opacity sources CH$_4$ and NH$_3$. 
The models shown here use line lists for CH$_4$ at $\lambda > 1.5 \mu$m, and CH$_4$ laboratory and 
astrophysical data (Strong et al. 1993; Karkoschka 1994) for $\lambda \leq 1.5 \mu$m. For NH$_3$, a line list is used for 
$\lambda > 1.4 \mu$m, but it is known to be  incomplete at 1.4 -- 1.9~$\mu$m and no 
NH$_3$ opacity is included at shorter wavelengths. Discrepancies between synthetic spectra and 
the observations can be seen in the $Y$-, $J$- and $H$-bands in Figure 4, most likely due to these incomplete opacities.  
Due to these  problems, we revisit the predictions of the SEDs of dwarfs cooler 
than 700~K, first from the observational and then from the theoretical point of view.

\subsection{Observational Indicators}

Figure 9 shows observed 0.96 -- 2.2~$\mu$m spectra of the 750~K dwarf 2MASS0415-09, the $\sim$160~K planet 
Jupiter{\footnote {from the IRTF Spectral Library, 
accessible at http://irtfweb.ifa.hawaii.edu/~spex/spexlibrary/IRTFlibrary.html, 
see Rayner, Cushing \& Vacca (2008)}},
and the $\sim$95~K planet Saturn.  Also shown are room-temperature laboratory data for CH$_4$ (Cruikshank \& 
Binder 1968) and NH$_3$ (Irwin et al. 1999). The physical environment of the brown dwarf atmosphere will be quite 
different from either the laboratory gas or the planets' atmospheres. Also, there is no H$_2$O in 
the planets' spectrum as all H$_2$O is condensed in deep cloud decks. Moreover, the planetary spectra are 
reflection spectra.  However,  both planets have NH$_3$ features in their spectra, with the warmer
Jupiter having less condensed NH$_3$ and stronger NH$_3$ gas absorption.  Given the 
absence of more appropriate data, we use these spectra to investigate what changes might occur at cold temperatures.

Starting at the longest wavelength, it can be seen that there is very little flux from Saturn or Jupiter in the 
$K$-band, due to strong pressure-induced H$_2$ opacity. Figure 3 shows that brown dwarfs also become very 
faint at $K$ with decreasing temperature. Figure 9 shows that the observed T8 $K$-band spectrum is very complex, due to the 
wealth of H$_2$O and CH$_4$, and possibly NH$_3$, absorption features 
(see also the higher resolution spectra of the T7.5 dwarfs in Figures 5 and 7, and the discussion of the Gl~229B 
$K$-band spectrum in Saumon et al. 2000).  The feature seen in 2MASS0415-09 at 2.02~$\mu$m is as strong 
in hotter 800~K spectra (Figures 1 and 3) and so is not a useful spectral indicator despite the 
apparent match to an NH$_3$ feature in the laboratory spectrum. Although the blue wing of this band 
was suggested by Burrows et al. (2003) as a region where increasing NH$_3$ absorption would become 
apparent for the temperature range of interest, Figures 3 and 9 show that 
identifying the increasing NH$_3$ opacity at 1.95 -- 2.05~$\mu$m amongst the existing features, 
and with low flux levels, is not practical. Hence the $K$-band is not an appropriate region to 
search for features that may indicate the next spectral type.

The $H$-band offers only a very narrow region of peak flux due to the strong H$_2$O features to the blue, 
and the strong CH$_4$ to the red of this band.  Identifying the increasing NH$_3$ opacity at 1.48 -- 
1.55~$\mu$m, as suggested by Burrows et al. (2003), again is problematical due to the already strong
H$_2$O absorption at these wavelengths.  
A steepening of the blue wing of the $H$-band peak may become visible, and we discuss this further 
below.   If the dip seen in the laboratory spectrum of NH$_3$ 
at 1.58~$\mu$m, which appears to carry through in the Jupiter
spectrum, is also seen in brown dwarf spectra, it may be a useful Y dwarf indicator, as it is in a region 
relatively free of H$_2$O and CH$_4$ absorption.

It appears that the $Y$ and $J$ regions may provide the best Y dwarf indicators.  This 
would be fortuitous, not only for the coincidence of the letter ``Y'', but because these 
extreme dwarfs emit more energy at 1.0--1.3~$\mu$m than at 1.5--2.2~$\mu$m.   
Weak absorption features in the 2MASS0415-09 spectrum at 1.21--1.22~$\mu$m and 1.29~$\mu$m  
apparently align with NH$_3$ features, and are weaker or non-existent in the warmer Gl~570D spectrum.
Even if these are due to CH$_4$ and not NH$_3$, if they become dominant features of colder dwarfs 
they could be used as spectral type indicators. The feature at 
1.27~$\mu$m is seen in the warmer spectrum and coincides with a strong sky emission feature, 
and so is not a useful spectral indicator. 

The blue wing of the $Y$-band peak may 
be a useful Y dwarf indicator due to the loss of gaseous K~I from the photosphere, as the strong 0.77~$\mu$m K~I 
feature impacts the blue edge of this band.
Also, the Jovian spectrum suggests that the 1.02--1.035~$\mu$m region may show NH$_3$ features.
The feature seen in 2MASS0415-09 at 1.02~$\mu$m is weak or non-existent in the 
Gl~570D spectrum, and so, whether due to NH$_3$ or CH$_4$, may be a useful later-T or Y dwarf indicator.

\subsection{Theoretical Indicators}

To further explore the appearance of NH$_3$ in very-late T dwarf spectra, we have complemented our NH$_3$ line opacity in 
the models with the Irwin et al. (1999) laboratory measurements, which were conducted at $T=230$ -- 296$\,$K at a resolution 
of 5$\,$cm$^{-1}$.  We applied their opacities over the range 0.91 -- 1.9$\,\mu$m by extrapolating the tabulated opacities 
to higher temperatures using their estimate for the energy of a representative lower level for the transitions in each 
frequency range.  As a test of the temperature extrapolation, we computed spectra with the measured opacities at a fixed 
temperature of 300$\,$K and obtained very similar synthetic spectra.

Figure 10 shows the resulting spectra for  brown dwarfs with $T_{\rm eff}$ of 600 and 700$\,$K.
A comparison of the 700$\,$K spectrum with one computed without the Irwin et al. (1999)
NH$_3$ opacity shows that a very strong feature would be expected on the blue side of the $H$ band
peak ($\sim 1.54\,\mu$m), that the $J$ band peak would be shaved on both sides, and that the blue
side of the $Y$ band peak would be suppressed by an NH$_3$ band.  While this would seem a promising way 
of identifying NH$_3$ features to define the spectral characteristics of Y dwarfs, none of these features
are apparent in the spectra of the coolest known T dwarfs with $T_{\rm eff} \sim 700$ -- 800$\,$K
(Figure 2).  The shoulders seen in the red wing of the $J$-band and the blue wing of the $H$-band are
in particular discord with the observational data (e.g. for 2MASS0415-09 in Figures 2, 4 and 9).

On the other hand, detailed analyses of cool T dwarfs have revealed a depletion of NH$_3$ in the atmosphere by a factor of 
$\sim 10$ due to non-equilibrium chemistry arising from vertical transport in the atmosphere (Saumon et al. 2006, 2007). 
Figure 10 shows a 600$\,$K spectrum with the Irwin et al. (1999) NH$_3$ opacity computed out of chemical equilibrium with an 
eddy diffusion coefficient of $K_{zz}=10^4\,$cm$^2\,$s$^{-1}$ (see e.g. Saumon et al. 2006). The NH$_3$ features from the 
Irwin et al. (1999) opacity are significantly weaker but remain stronger than seen in the 
observational data.  It is highly
unlikely that non-equilibrium chemistry can further decrease the NH$_3$ abundance by the required factor.  Additional 
laboratory work on the opacity of NH$_3$ well above room temperature, as well as {\it ab initio} line list calculations,
are urgently needed to understand the spectra of the new cooler brown dwarfs that will soon be uncovered by modern large 
surveys.

\section{Conclusions}

We have used the well-studied T7.5 and T8 dwarfs Gl~570D and 2MASS0415-09 as templates to determine 
physical parameters for the very similar dwarfs HD~3651B, 2MASS09393-24 and 2MASS1114-26.  Synthetic 
spectra and photometry, together with new and published near-infrared spectra and mid-infrared photometry, 
show that HD~3651B, a distant companion to an exoplanet host star, has both higher gravity and higher 
metallicity than Gl~570D.  The derived parameters for this dwarf are: $T_{\rm eff}=$820 -- 830~K, 
log~$g=$5.4 -- 5.5, [m/H]$=+$0.2 and $K_{zz}=10^4$~cm$^2$s$^{-1}$.  The corresponding age is around 10~Gyr 
and mass 65~M$_{\rm Jupiter}$. Synthetic spectra, together with published near-infrared spectra, show that 
2MASS0939-24 and 2MASS1114-26 are both metal-poor compared to 2MASS0415-09, and have similar or slightly 
higher gravities.  The derived parameters for these dwarfs are: $T_{\rm eff}=$725 -- 775~K and 
[m/H]$=-$0.3; log~$g=$5.3 -- 5.45 for 2MASS0939-24 and log~$g=$5.0 -- 5.3 for 2MASS1114-26. The 
corresponding ages and masses are $\sim$10~Gyr and 60~M$_{\rm Jupiter}$ for 2MASS0939-24, and $\sim$5~Gyr 
and 40~M$_{\rm Jupiter}$ for 2MASS1114-26. These temperatures are significantly higher than those derived 
by BBK, and hence these objects are not the coolest known T dwarfs. For these two dwarfs (which do not have
published parallax values) the atmospheric 
properties are mostly derived from the ratio of the $J$ and $K$ flux peaks, which is sensitive to both 
gravity and metallicity, and the shape of the $Y$ flux peak, which is sensitive primarily to metallicity.

The intrinsic accuracy of the inferred atmospheric parameters depends, of course, on the veracity of the 
atmosphere models.  Until other model sets are similarly applied to comparable datasets, the systematic 
uncertainties are difficult to evaluate.  However, given the apparent simplicity of late T dwarfs 
(relative to the cloudy L dwarfs, for example), large systematic 
errors are unlikely.  Errors larger than 5 to 10\% in effective temperature and 0.3 dex in $\log g$ and 
metallicity would be surprising.  The uncertainties here in $\log g$ and metallicity are smaller than this 
and are estimated to be 0.1~dex (or 26\%), as we  have carried out relative comparisons to the well-studied 
dwarfs Gl~570D and 2MASS0415-09. 

When the metallicities and gravities of more T dwarfs have been determined (following 
improvements to current models), the community will need to agree on metallicity and gravity labels in
spectral classification schemes.  We suggest that these labels should follow existing stellar schemes.
Metal-poor main-sequence M dwarfs are labelled as ``subdwarfs'' due to their location in optical
color-magnitude diagrams (e.g. Figure 10 of Monet et al. 1992).  Although the location of brown dwarfs in 
such diagrams will be a strong function of gravity, the ``sd'' prefix could be maintained as a 
metal-poor indicator, as already being done for some L dwarfs (e.g. Burgasser 2004).
Similarly, gravities should be described numerically, using the log of the value in 
c.g.s. units, following the spectral type. It would be desirable to specify gravity to at least 
one-half a dex, as this parameter has a large impact on brown dwarf SEDs. Also, for a given 
$T_{\rm eff}$, a range of 1.0~dex in log~$g$ implies a large range in age and mass
(for example if $T_{\rm eff}= 800$~K, log~$g=4.5$ corresponds to mass and age around 15~M$_{\rm Jupiter}$
and 0.5~Gyr, while log~$g=5.5$ corresponds to 70~M$_{\rm Jupiter}$ and 15~Gyr).  
Given this requirement,
and the existing association of roman numerals with luminosity class, we suggest that arabic numerals are
used for brown dwarf gravities, following a ``g'' to separate it from the spectral type.  
Thus a metal-poor T7.5 dwarf with log~$g=5.5$ would be described as sdT7.5g5.5. For L and early-T dwarfs
an additional classification dimension will be required to describe the ``dustiness'' or ``redness'' of
the spectrum caused by the condensate cloud decks, perhaps using the letter ``r'' 
(as ``d'' would be confused with ``dwarf''). 

Existing models would  greatly benefit from improved line lists for 
transitions of CH$_4$ and NH$_3$ at $\lambda \leq 1.7 \mu$m.  Such lists would  allow substantially better 
analyses of the SEDs of the T dwarfs, and would yield better predictions for the 
spectra and colors of the next, cooler, spectral type. 
We use observed near-infrared spectra of 2MASS0415-09, Jupiter and Saturn, and laboratory observations of 
CH$_4$ and NH$_3$ gas, to investigate the appearance of this spectral type.  The expectation was 
that the appearance of NH$_3$ features in the blue wings of the $H$ and $K$ flux peaks would be significant
enough to require a new (``Y'') spectral type, and that this would occur at $T_{\rm eff}\approx 600$~K.
However, we find that the lack of flux in the $H$ and $K$ wings,
combined with the complexity of already strong H$_2$O and CH$_4$ absorption features, make it unlikely that these 
NH$_3$ features will be useful.  Instead we find that previously overlooked NH$_3$ features 
that occur in the brighter $Y$- and $J$-bands
may  be the indicator of the next spectral type, combined with the changing shape of the $Y$-band 
as K~I solidifies into KCl. Large uncertainties remain still, as 
an attempt to include NH$_3$ laboratory data into our models produced spectra that do not agree 
with observations, and as it is now known from mid-infrared analyses that 
the abundance of NH$_3$ in late-T dwarfs is around 10 times lower than expected from chemical equilibrium models.
It is possible that the next spectral type may not be triggered until much cooler temperatures are reached. 
Progress will be made as observers
continue to search to fainter limits to find cooler, and possibly spectrally different, brown dwarfs,
while theoreticians continue  improving
the opacity line lists and models.

\acknowledgments

We are grateful for the award of Director's Discretionary time at the Gemini
Observatory, which is operated by the
Association of Universities for Research in Astronomy, Inc., under a cooperative agreement
with the NSF on behalf of the Gemini partnership: the National Science Foundation (United
States), the Particle Physics and Astronomy Research Council (United Kingdom), the
National Research Council (Canada), CONICYT (Chile), the Australian Research Council
(Australia), CNPq (Brazil) and CONICET (Argentina).
TRG and SKL are supported by the Gemini Observatory, and acknowledge the above partnership
for this support. This work was supported in part under the auspices of the U.S. Department of
Energy at Los Alamos National Laboratory under Contract W-7405-ENG-36.
MSM acknowledges the support of the NASA Office of Space Sciences. 
MCL acknowledges support for this work from NSF grants AST-0407441 and
 AST-0507833 and an Alfred P. Sloan Research Fellowship.
Dale Cruikshank gave 
invaluable assistance with the CH$_4$ and NH$_3$ laboratory data. Katarina Lodders contributed
significantly in many helpful discussions. John Rayner generously provided the Jovian spectrum pre-publication.
Finally, we are grateful to the referee, who significantly improved the manuscript.


\begin{figure} \includegraphics[angle=-90,scale=.65]{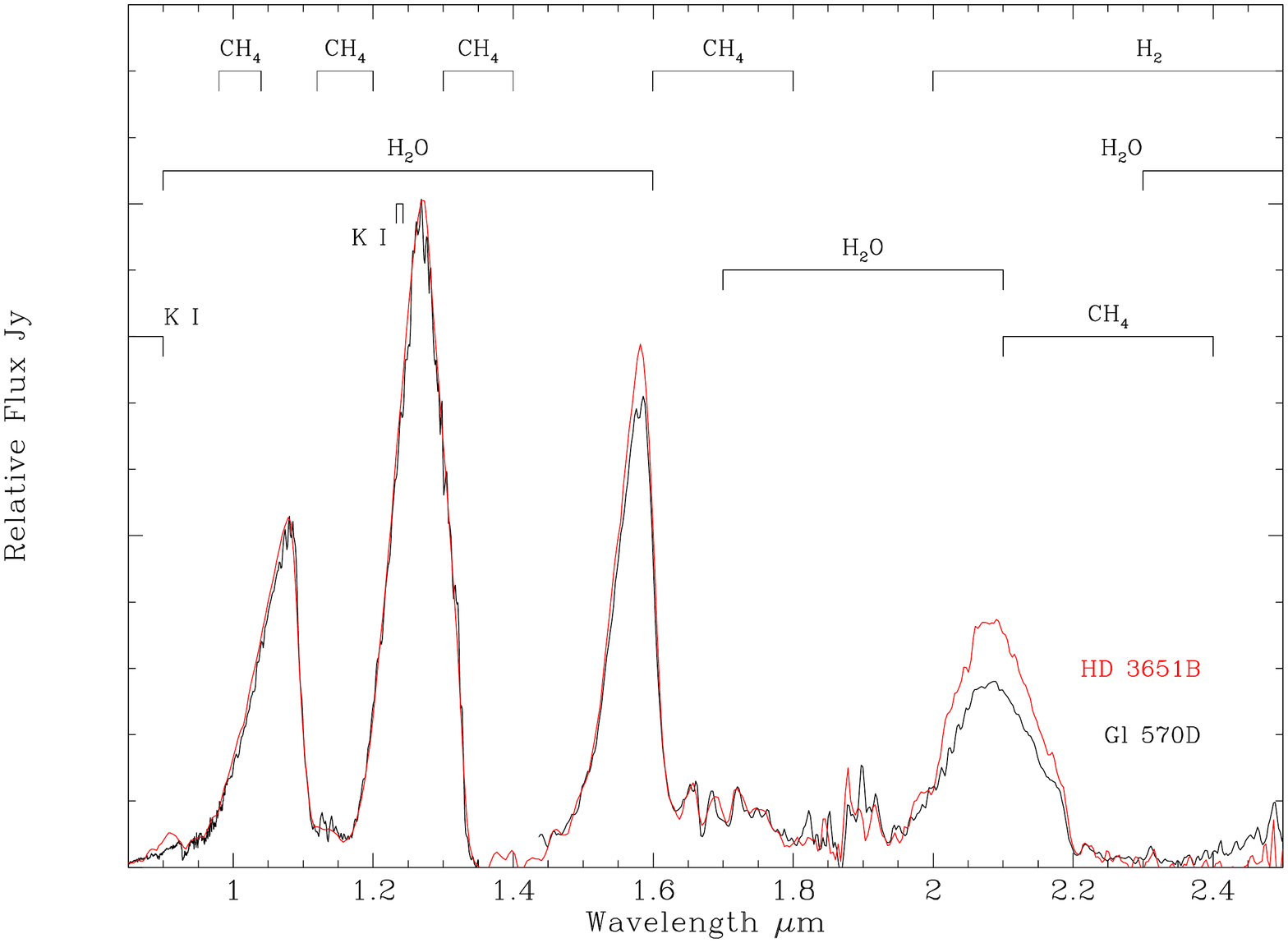} 
\caption{Observed R$\approx$400 CGS4 near-infrared spectra of Gl~570D 
(Geballe et al. 2001) and  R$\approx$150 Spex spectra of HD~3651B (Liu et al. 2007).  
Spectra are scaled to the peak flux at 1.27~$\mu$m.}
\end{figure}
\clearpage

\begin{figure} \includegraphics[angle=-90,scale=.65]{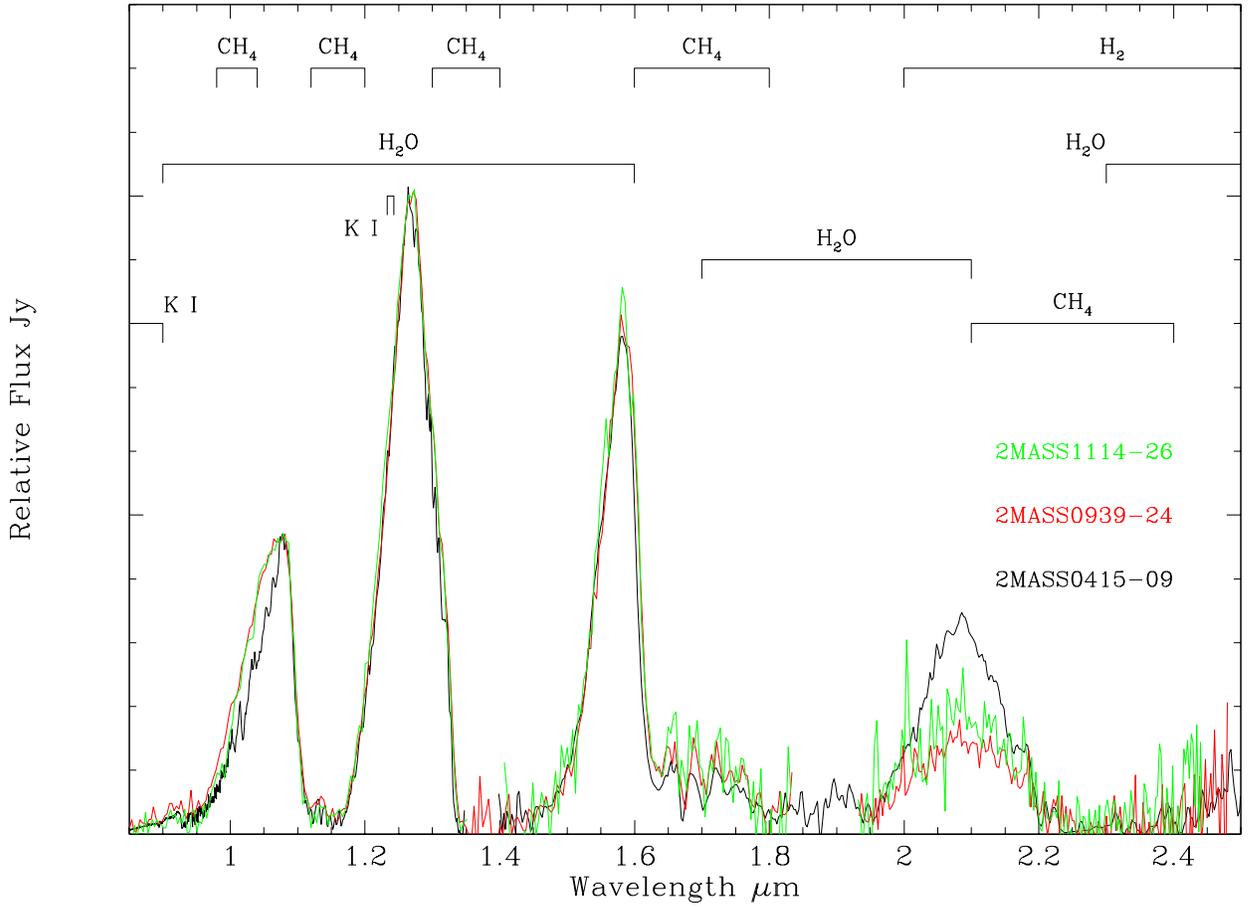} 
\caption{Observed R$\approx$400 CGS4 near-infrared spectra of 2MASS0415-09
(Knapp et al. 2004) and 
R$\approx$150 Spex spectra of 2MASS0939-24 and 2MASS1114-26
(Burgasser, Burrows \& Kirkpatrick 2006b).  
Spectra are scaled to the peak flux at 1.27~$\mu$m. }
\end{figure}
\clearpage

\begin{figure} \includegraphics[angle=0,scale=.70]{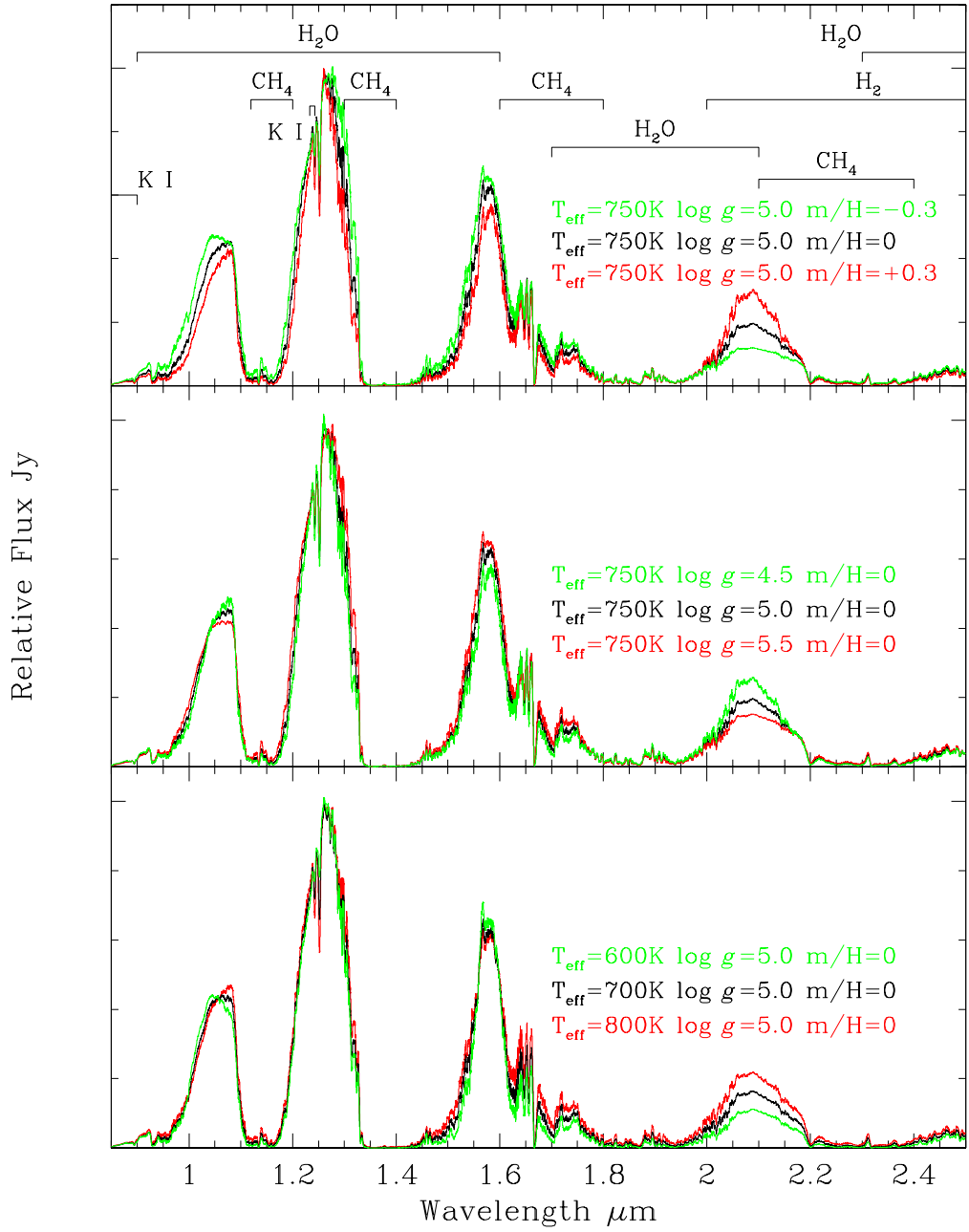} 
\caption{Synthetic spectra smoothed to R$=$400 and scaled to the peak flux at 1.27~$\mu$m.
}
\end{figure}
\clearpage

\begin{figure} \includegraphics[angle=0,scale=.70]{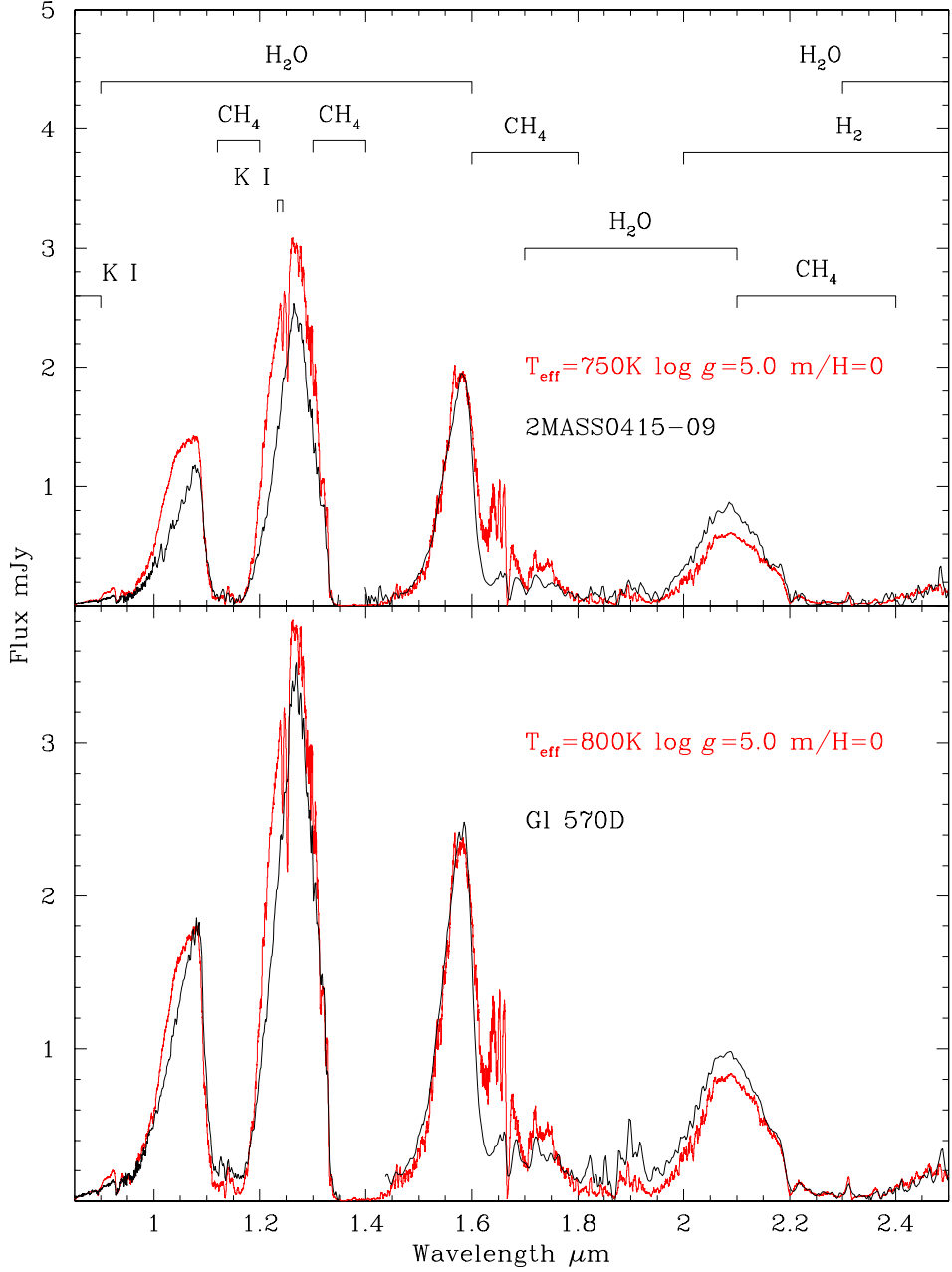} 
\caption{Observed R$\approx$400 near-infrared spectra of Gl~570D and 2MASS0415-09 (black),
and synthetic spectra scaled for the observed flux at Earth (red).
}
\end{figure}
\clearpage

\begin{figure} \includegraphics[angle=0,scale=.70]{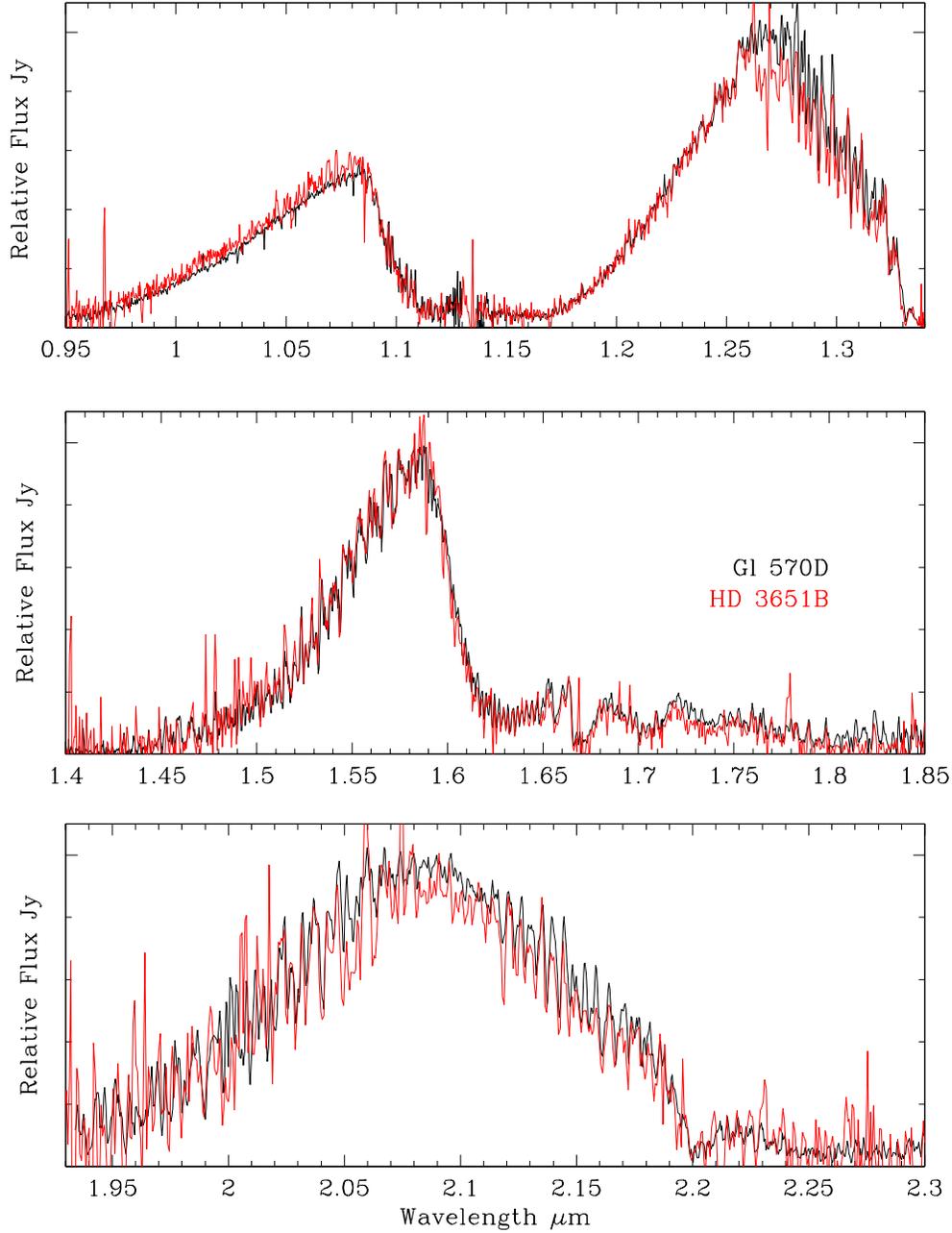} 
\caption{Observed medium-resolution R$=$1700--2000 near-infrared spectra of Gl~570D and HD~3651B, 
from the NIRSPEC brown dwarf spectroscopic survey (McLean et al. 2003) and this work. 
Spectra are scaled to the peak flux in each panel. 
See Figure 1 for line identifications.}
\end{figure}
\clearpage

\begin{figure} \includegraphics[angle=-90,scale=.60]{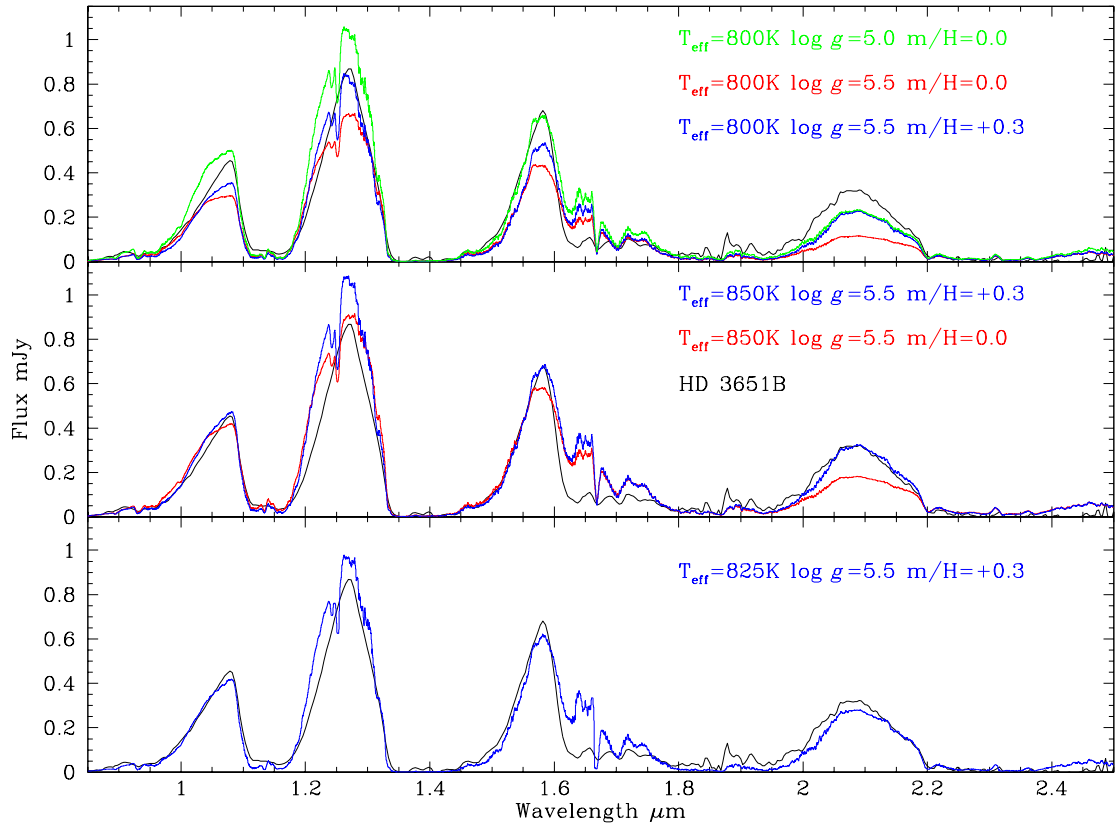}
\caption{Observed low-resolution near-infrared spectrum of HD~3651B (black curve), compared to synthetic 
spectra scaled for the observed flux at Earth for various temperatures, gravities and metallicities.}
\end{figure}
\clearpage

\begin{figure} \includegraphics[angle=-90,scale=.60]{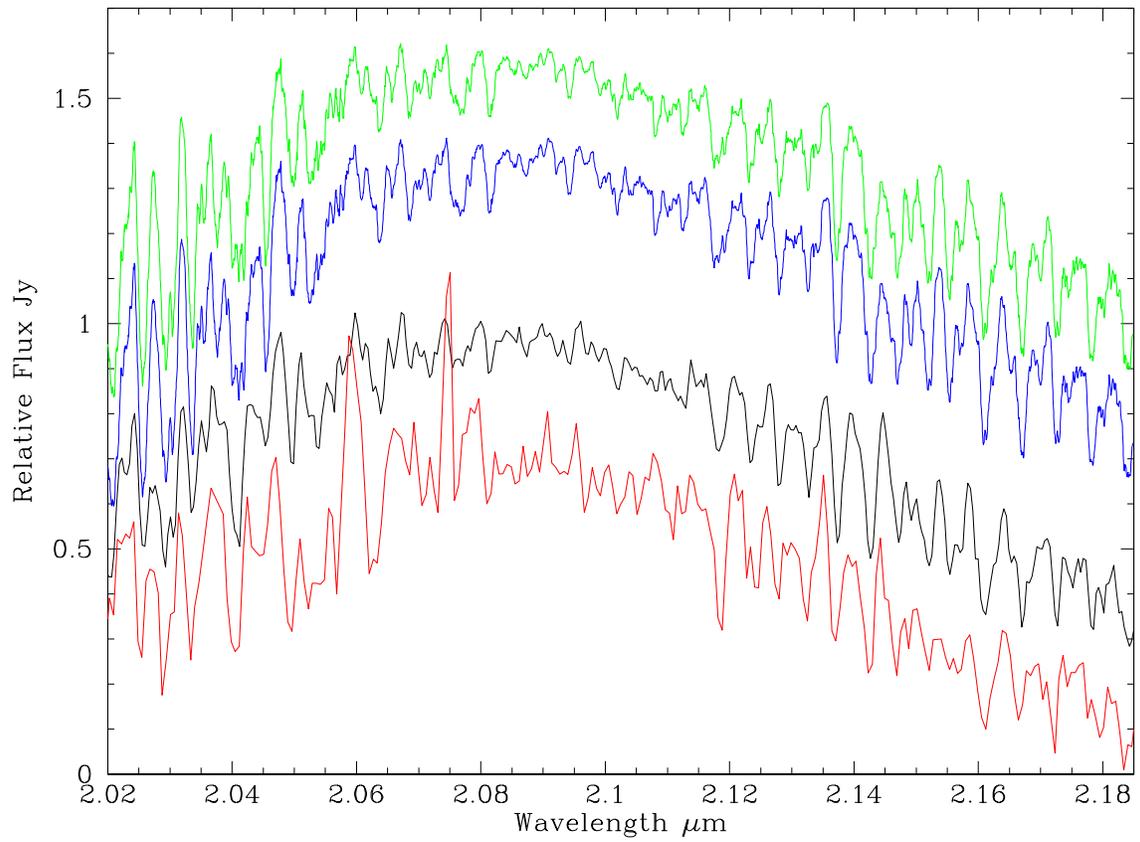}
\caption{Medium-resolution spectra of Gl~570D (McLean et al. 2003, black) and HD~3651B (this work, red)
compared to synthetic spectra for [$T_{\rm eff}$, log~$g$, [m/H]] models [800, 5.0, 0.0] (green) and [800, 5.5, +0.3] (blue). The spectra are scaled to the peak flux and offset for clarity.
}
\end{figure}
\clearpage

\begin{figure} \includegraphics[angle=-90,scale=.60]{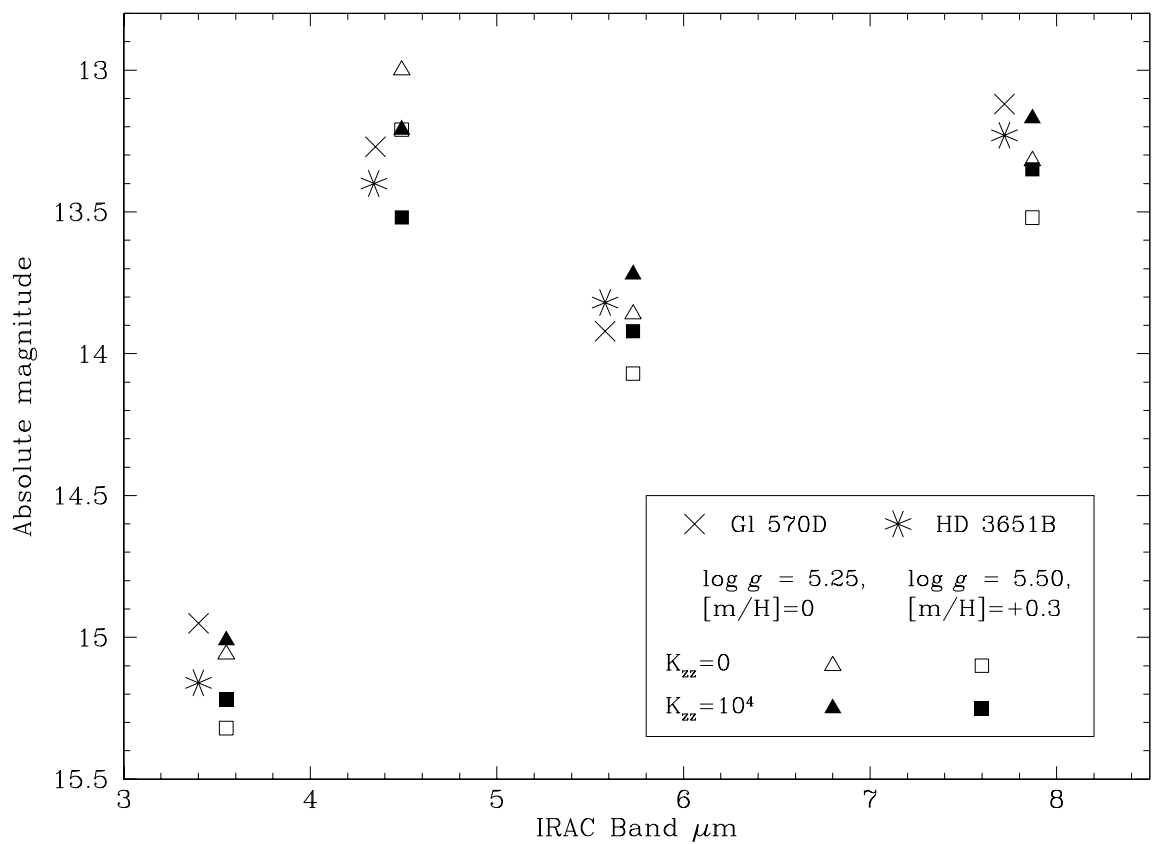}
\caption{Modelled (for $T_{\rm eff}=825$~K)
and observed absolute IRAC magnitudes, symbols are described in the legend. 
Observations are taken from Patten et al. (2006) and Luhman et al. (2007); errors in the photometry
are 2--5\% for both dwarfs for [3.55] and [4.49], and 7--14\% at [5.73] and [7.87].
Observational datapoints are shifted to shorter wavelengths for clarity.}
\end{figure}
\clearpage

\begin{figure} \includegraphics[angle=-90,scale=.60]{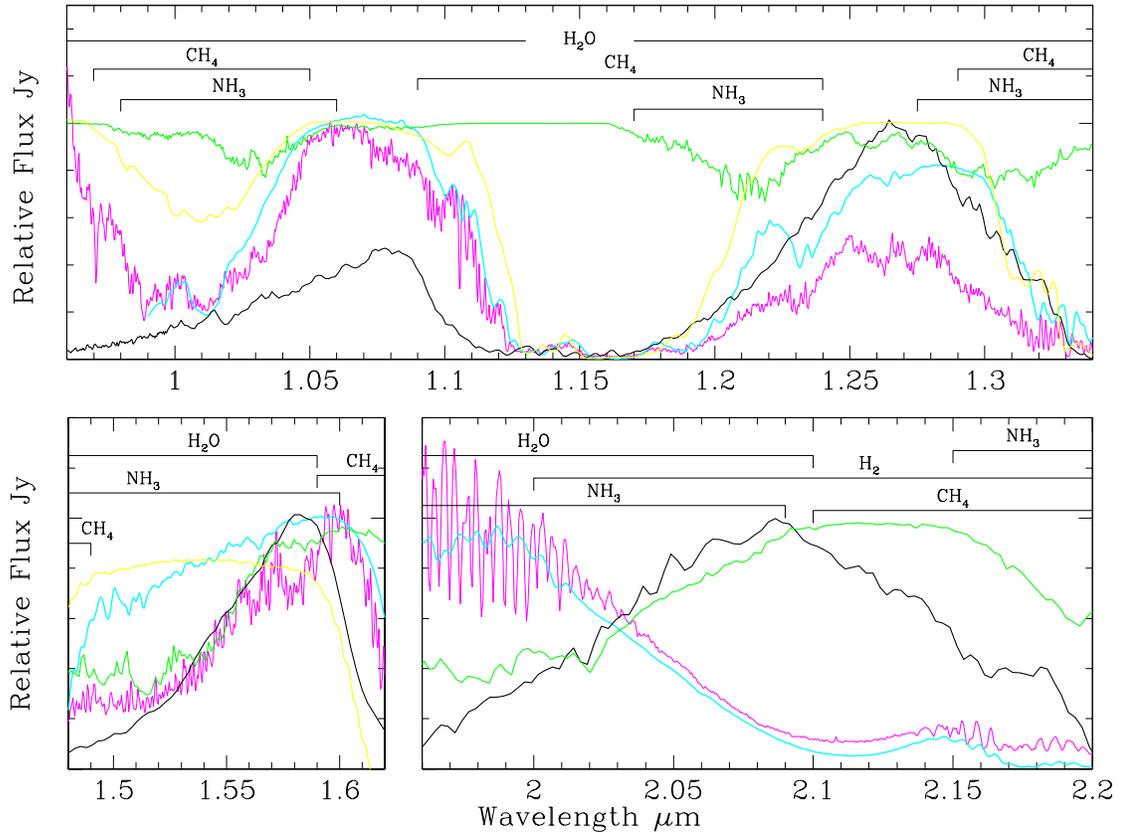} 
\caption{ Spectra of the T8 dwarf 2MASS0415-09 (black, Knapp et al. 2004), Jupiter (magenta, 
Rayner, Cushing \& Vacca 2008) and Saturn (cyan, Kim 
\& Geballe 2005), together with room temperature transmission laboratory data for CH$_4$ (yellow, Cruikshank 
\& Binder 1968) and NH$_3$ (green, Irwin et al. 1999). } \end{figure} \clearpage

\begin{figure} \includegraphics[angle=-90,scale=.60]{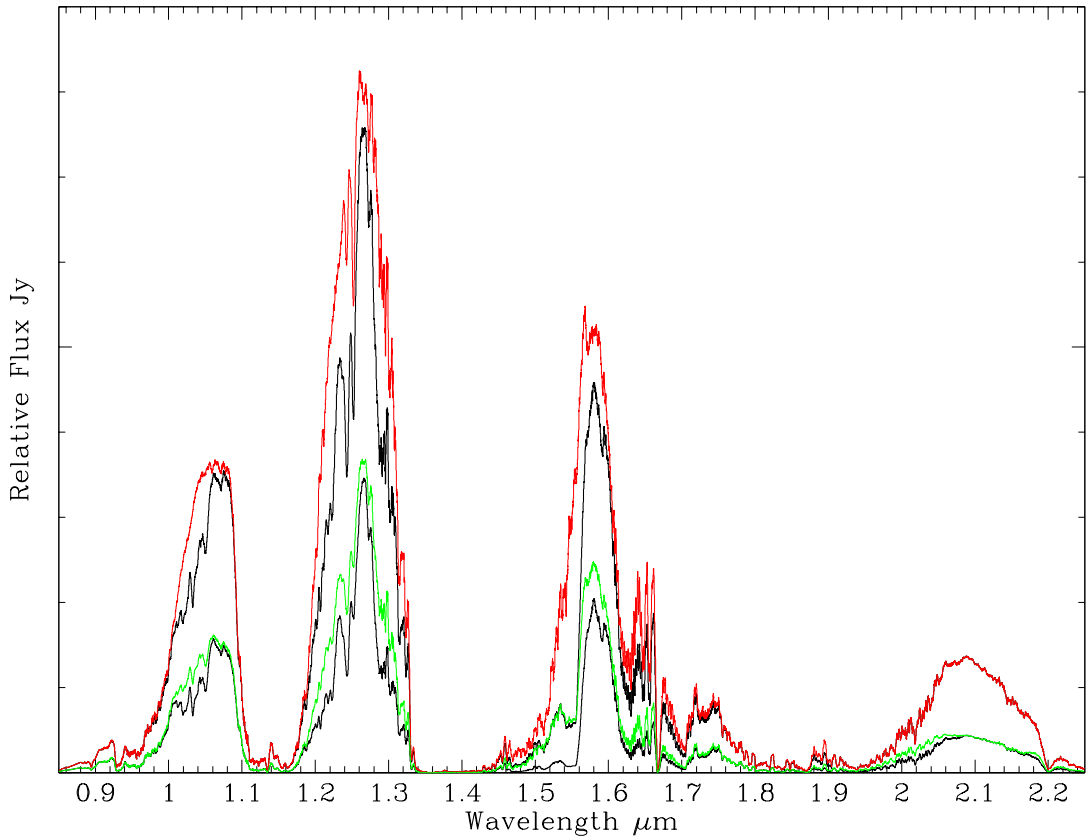}
\caption{
Ammonia features in the near-infrared spectra of very cool brown dwarfs.
         The black curves show cloudless synthetic spectra with $T_{\rm eff}=700\,$K 
         (upper curve) and 600$\,$K (lower curve), $\log g=5$, and including the
         NH$_3$ opacity of Irwin et al. (1999) for $\lambda < 1.9\,\mu$m and the
         NH$_3$ opacity computed from a line list for $\lambda > 1.9\,\mu$m.  The upper 
         curve (red) shows the 700$\,$K model computed without the Irwin et al. opacity.
         A 600$\,$K model computed with the Irwin et al. 
         opacity but with NH$_3$ depleted by non-equilibrium chemistry 
         (with the eddy diffusion coefficient 
         $K_{zz}=10^4\,$cm$^2$/s) is shown by the lower curve (green). All spectra for a
         given $T_{\rm eff}$ are computed with the same $(T,P)$ structure. The spectra 
         have been smoothed to a resolving power of R=400.}
\end{figure}
\clearpage

\voffset 1.5truein
\begin{deluxetable}{llrrrccccccccc}
\rotate
\tabletypesize{\scriptsize}
\tablecaption{Published Physical Parameters for  T7.5--T8 Dwarfs}
\tablewidth{625pt}
\tablehead{
\colhead{Name} &  \colhead{Spectral} & \colhead{Parallax}  & \colhead{$V_{\rm tan}$(error)} &   
\colhead{log($L_{\rm bol}/$} & &\multicolumn{4}{c}{Parameters from Luminosity} & &
\multicolumn{2}{c}{BBK Parameters\tablenotemark{b}} 
&  \colhead{References\tablenotemark{c}} \\
\cline{7-10}\cline{12-13} 
&\colhead{Type\tablenotemark{a}} & \colhead{(error)(mas)}& \colhead{(km~s$^{-1}$)} & 
 \colhead{$L_{\odot}$)(error)} & &
\colhead{$T_{\rm eff}$~K} & \colhead{log~$g$} & \colhead{Age Gyr} & \colhead{Mass M$_J$} & &
\colhead{$T_{\rm eff}$~K} & \colhead{log~$g$} & 
\\
}
\startdata
2MASS J04151954-0935066         & T8   & 174.34(2.76) & 61.4(1.0) & -5.67(0.02) & &
725--775 & 5.00--5.37 & 3--10 & 33--58 && 740--760  & 4.9--5.0 & 1, 2, 3  \\
2MASS J09393548-2448279         & T8   & \nodata      & 46(8)   & \nodata & &
\nodata & \nodata &   \nodata & \nodata & & $\lesssim$ 700 &   \nodata &  4 \\
2MASS J11145133-2618235         & T7.5 & \nodata      & 109(20)   & \nodata & &
\nodata & \nodata &   \nodata & \nodata & & $\lesssim$ 700 &   \nodata & 4 \\
2MASS J12171110-0311131         & T7.5 & 93.2(2.06)   & 53.5(1.2) & -5.31(0.03) & &
850-950 & 4.80--5.42 & 1--10 & 25--66 & & 860--880 & 4.7--4.9 & 2, 3, 5, 6  \\
Gl 570 D & T7.5 & 170.16(1.45) & 56.4(0.9)   & -5.54(0.01) & &
800-820 & 5.09--5.23 & 3--5 & 38--47 && \nodata & \nodata & 7 , 8, 9, 10\\
HD 3651B & T7.5 & 90.03(0.72) & 31.0(1.2)    & -5.60(0.05) & &
780--840 & 5.1--5.5 & 3--12 & 63--72 & & 760--820 &  4.7--5.3 & 11, 12, 13, 14 \\
\enddata

\tablenotetext{a}{~Spectral types are based on the near-infrared classification scheme for T dwarfs 
by Burgasser et al.\ (2006a) and have an uncertainty of 0.5 of a subclass. }
\tablenotetext{b}{~Derived from the near-infrared spectral ratio technique of 
Burgasser, Burrows \& Kirkpatrick (2006b)}
\tablenotetext{c}{
(1) Burgasser et al. (2002); (2) Vrba et al. (2004); (3) Saumon et al. (2007);
(4) Tinney et al. (2005),  
tangential velocities are estimated from distances derived from $M_J$:spectral type;
(5) Burgasser et al. (1999); (6) Tinney et al. (2003); 
(7) Burgasser et al. (2000);  (8) ESA (1997) 
(9) van Altena, Lee \& Hoffleit (1995) (10) Saumon et al. (2006) 
(11) Mugrauer et al. (2006) (12) Luhman et al. (2007) (13) Burgasser (2007) (14) Liu et al. (2007)
}

\end{deluxetable}
\clearpage

\begin{deluxetable}{llccccc}
\tabletypesize{\scriptsize}
\tablecaption{Derived Properties of HD~3651B, 2MASS J09393548-2448279 and 2MASS J11145133-2618235}
\tablewidth{400pt}
\tablehead{
\colhead{Name} &  \colhead{Spectral} & \colhead{Metallicity}  & 
\colhead{$T_{\rm eff}$~K} & \colhead{log~$g$\tablenotemark{c}} & \colhead{Age Gyr\tablenotemark{c}} & \colhead{Mass M$_J$\tablenotemark{c}} \\
&\colhead{Type\tablenotemark{a}} & \colhead{[m/H]\tablenotemark{b}}&  &  & & \\ 
}
\startdata
HD~3651B &   T7.5 & $+$0.2 & 820--830 & 5.4--5.5 & 8--12 & 60--70 \\
2MASS J09393548-2448279 & T8 & $-$0.3 & 725-775 & 5.3--5.45 & 10--12 & 50--65 \\
2MASS J11145133-2618235 & T7.5 & $-$0.3 & 725-775 & 5.0--5.3 & 3--8 & 30--50 \\
\enddata
\tablenotetext{a}{Uncertainty in spectral type is 0.5 of a subclass.}
\tablenotetext{b}{Uncertainty in metallicity is estimated to be 0.1 dex.}
\tablenotetext{c}{Gravity, age and mass are correlated such that lower gravity 
implies younger age and lower mass, and higher gravity implies an older and more massive dwarf.}
\end{deluxetable}

\end{document}